# LIGO series, dimension of embedding and Kolmogorov's complexity.

Marcelo G. Kovalsky and Alejandro A. Hnilo
*CEILAP, Centro de Investigaciones en Láseres y Aplicaciones, (MINDEF-CONICET);*
*J.B. de La Salle 4397, (1603) Villa Martelli, Argentina.*
*email: ahnilo@citedef.gob.ar*
August 10th, 2020.

The interpretation of the series recorded by the Laser Interferometer Gravitational Wave Observatory is a very important issue. Naturally, it is not free of controversy. Here we apply two methods widely used in the study of nonlinear dynamical systems, namely, the calculation of Takens' dimension of embedding and the spectrum of Kolmogorov's complexity, to the series recorded in event GW150914. An increase of the former and a drop of the latter are observed, consistent with the claimed appearance of a gravitational wave. We propose these methods as additional tools to help identifying signals of cosmological interest.

The first detection of gravitational waves by the LIGO (Laser Interferometer Gravitational-Wave Observatory) project is one of the major achievements in Physics in this century. The cosmological signals are predicted to be tiny, and must be rescued from a sea of noise produced by many different sources. In few words, the essence of the method to identify gravitational waves is to compare the raw time series recorded by the interferometers with a set of "templates". These templates correspond to signals produced by probable cosmological sources, calculated according to the theory of General Relativity. However, in a long series, it cannot be discarded these templates to be produced, just by chance, by the noise. It is then crucial to determine the coincidence (at the right timing) of template-fitting signals observed at distant stations. These signals, after corrections due to the different orientation of the interferometers, reveal information on the features of the cosmological source.

A key point is that the residual series, obtained by subtracting the templates from the raw signals, should show no correlation among them. Here is where controversy arises, because the residual series are observed not to be, in general, fully uncorrelated [1]. On the one hand, some remaining correlation is to be expected for no numerical template can fit an actual signal exactly. On the other hand, the remaining correlation leads to methodological doubts on the cause of the observed signals.

We do not take sides on the controversy. The aim of this paper is to call the attention to a couple of well tested numerical tools that, used in combination with the already established methods, may provide additional information to interpret the data. These tools are the dimension of embedding dE [2] and Kolmogorov's complexity Kc [3-5]. A change in dE may reveal the emergence of a new object in phase space, regardless its detailed features. A drop in Kc may reveal the emergence of a component describable by an equation, regardless the equation. Both tools seem especially suitable to help to analyze the series recorded at LIGO, for the meaning of their results is independent of the correlation between the residual series.

We calculate dE and the spectra of Kc for the series in event GW150914 (see Figure 1), which is the claimed first reported detection of a gravitational wave. Our results indicate these series to reveal the emergence of a distinct object of low complexity in phase space, which has the same dimension in both stations, and which is not present in the residual series. These results are consistent with the detection of a gravitational wave.

In what follows, the meaning of dE and Kc are briefly described. After that, the obtained results are discussed in detail.

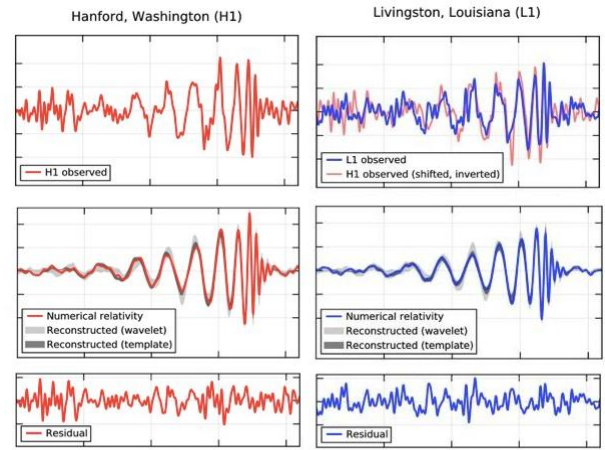

Figure 1: Recorded series of event GW150914. Left column (red) was recorded at Hanford, right column (blue) at Livingston, propagation delay is taken into account, horizontal scale 50 ms/div. Upper row corresponds to the two raw series, middle row to the theoretical templates, row in the bottom to what is left after subtracting the templates from the raw series (adapted from www.gw-openscience.org/events/GW15091).

The dimension of embedding dE is a concept from nonlinear analysis. In a chaotic system the evolution can be apparently random. Nevertheless, it involves few degrees of freedom linked through nonlinear equations, and is even partially predictable (up to some horizon of predictability). Takens' reconstruction theorem allows measuring the number of dimensions dE of the object in phase space within which the system (that generates the series in question) evolves. The method is based in measuring the decay of the number of false nearest neighbors (*fnn*) as the number of dimensions of the reconstruction increases [2]. The reconstruction is reliable as far as the number of *fnn* is sufficiently large. Software devised for this purpose control this condition to hold. Here we use the program codes in the freely available TISEAN set, so that our results can be checked by any interested researcher.

The complexity Kc of a binary series of length *N* is the binary length of the shortest program (running on a classical Turing machine) whose output is the said series. As there is no way of expressing the series using less bits that the series itself, the series is said to be *incompressible*. This definition is intuitive and appealing, but has the drawback that Kc cannot be properly computed. For, one can never be sure that there is no shorter program able to generate the series. Nevertheless, Kc can be *estimated* from the rate of compression as the coding alphabet increases using, f.ex., the algorithm devised by Lempel and Ziv [6]. Here we use the approach developed by Kaspar and Schuster [7] and implemented by Mihailovic [8] to estimate a normalized complexity Kc. This value is designed to be near to 0 for a periodic or regular sequence, and near to 1 for a fully incompressible one. Complexity has advantages over other methods of detecting regular behavior: it does not need the series to be stationary, and applies to series of any length. On the other hand, in addition to its intrinsic non-computability, the value of Kc depends on the amplitude threshold used to compose a binary series from the raw data. That's why it has been recently proposed the use of a *spectrum of complexity* [8] obtained by plotting the values of Kc calculated for each possible threshold value, see Figure 2. The maximum of this spectrum provides a reliable evaluation of complexity, for it is independent of the threshold. Besides, the area under the curve defines the *total* complexity.

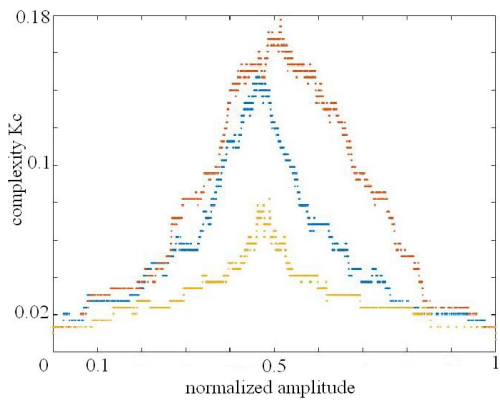

Figure 2: Spectrum of complexity of the series recorded at Hanford. Orange: residual, blue: raw, yellow: template. The curves for the series recorded at Livingston are similar.

We have successfully applied dE and Kc to the experimental study of nonlinear dynamics of self-pulsing lasers [9], to search deviations from ergodicity in Bell's experiments [10], and to evaluate randomness of series of numbers generated using entangled states of photons [11]. In general, we observed "noise" to be related with the large number of degrees of freedom of the environment. Noise had, in consequence, no measurable value of dE. For the same reason, it was incompressible and had a high value of Kc. "Signal", on the other hand, was determined by some underlying cause describable by an equation of evolution, or algorithm. It usually had a measurable value of dE and relatively low Kc.

In the case of LIGO signals instead, we are surprised by the dynamical simplicity of both signal and noise. Both have well defined quasi-periodical features (see Fig.1). The ideal signal (template) is a chirped wave packet with frequencies around 120 Hz. The first minimum in the mutual information, which is the standard chosen delay to reconstruct objects in phase space [1], is found to be 27 for residual series of 3441 elements. This means the presence of a strong self-correlation along these series (see also below).

In consequence, a low value of dE is expected, and in fact obtained with reliable values of *fnn*, for both the theoretical template and the experimentally obtained series. Yet, the raw (signal + noise) series have a larger value of dE. This implies they are the composition of two different objects in phase space, one corresponding to the quasi-periodical noise, the other one to the (presumed) gravitational wave.

As it can also be expected, the values of Kc for the templates (which are generated by an algorithm) are much smaller than for the residual noises. This is natural because noise, even if it is mainly quasi-periodical here, necessarily includes some degree of non-predictability. The complete (or raw) series have an intermediate value of Kc. The precise results for the series in GW150914 are displayed in Table 1. The spectra of complexity (see Fig.2) clearly distinguish the three cases.

|  | *Raw* | *Residual* | *Template* |
|---|---|---|---|
| *Kc max ( H1)* | 0.15 | 0.175 | 0.08 |
| *dE (H1)* | 3 | 2 | 2 |
| *Kc max ( L1)* | 0.13 | 0.16 | 0.075 |
| *dE (L1)* | 3 | 2 | 2 |

Table 1: Values of complexity (maximum of the spectrum) and dimension of embedding for the time series recorded at Hanford (H1) and Livingston (L1).

The difficulties in dealing with noise in this experiment are known to be formidable. As an additional element to consider in this sense, the Hurst coefficient (which measures self-correlation decay) in the residual series is 0.548 for Hanford and 0.664 for Livingston. These numbers indicate there is a long range memory effect, or persistence, in the residual noise. This unusual feature is perhaps at the root of the mentioned controversy. Besides, standard Augmented Dickey-Fuller (ADF) and Kwiatkowski–Phillips–Schmidt–Shin (KPSS) tests show these series to be non-stationary. This result confirms that their statistical features change rapidly, and hence, that they are impossible to estimate and to subtract to recover the signal. Finally, surrogated series have the same values of dE and Kc than the actual ones. This means that, in spite of the quasi-periodical feature of the residual noise, there is no underlying dynamical cause that can be used to help to filter it out.

In summary: the raw series produced in event GW150914 show an increase in the value of dE and a drop in the maximum of Kc spectrum in comparison with the residual ones. These results are consistent with the emergence of a distinct object of low dimension and low complexity in phase space. This event occurs in both stations with the delay given by the velocity of light, shows the same value of dE and a comparable drop in Kc. It is then reasonable to infer the emerging object to be produced by a gravitational wave. The Hurst, KPSS and ADF coefficients of the residual series indicate that they are an unusual kind of noise. It is statistically non-stationary and displays long range correlations. These features may explain the origin of the controversy.

**Acknowledgments.**

This material is based upon research supported by, or in part by, the U. S. Office of Naval Research under award number N62909-18-1-2021. Also, by grant PIP 2017-027 CONICET (Argentina).